# Frequency perturbation integral for piezoelectric quartz crystal microbalances based on scalar differential equations

Jiashi Yang (jyang1@unl.edu)
Department of Mechanical and Materials Engineering
University of Nebraska-Lincoln, Lincoln, NE 68588-0526, USA

*Abstract* – We study frequency shifts in a piezoelectric quartz resonator induced by a surface mass layer for sensor applications. The scalar differential equations for thickness-shear modes in a quartz plate are used. A first-order perturbation analysis is performed. The frequency shift is obtained and is expressed by a perturbation integral. It produces the well-known Sauerbrey equation for mass sensitivity in the special case of a uniform mass layer. As an application of the perturbation integral, frequency shifts due to a nonuniform mass layer are calculated.

## I. INTRODUCTION

Piezoelectric quartz crystals have been used for a long time to make acoustic wave resonators as frequency standards for timing and frequency control. Because of material anisotropy and electromechanical couplings, theoretical and numerical modeling of piezoelectric resonators present considerable mathematical challenges. For the basic operating thickness-shear mode of the most widely used AT-cut quartz resonators, Tiersten derived a two-dimensional (2-D) scalar differential equation [1-4] from the three-dimensional equations of piezoelectricity. The scalar equation was later generalized to quartz resonators of doubly-rotated cuts with general material anisotropy and more operating modes [5-7]. During the past few decades, these scalar equations were used extensively by many researchers to obtain the resonant frequencies, modes and capacitance of various electroded or unelectroded quartz resonators of rectangular, circular and elliptical geometry, contoured resonators with nonuniform thickness, and filters with two or more pairs of electrodes [8-41].

Relatively recently, quartz resonators have been used to make mass sensors called quartz crystal microbalances (QCMs) based on the frequency effect of a thin mass layer on a resonator. For modeling QCMs, the scalar equations were generalized to include the mechanical effects of surface mass layers in [42, 43] through surface acoustical impedance.

In this paper we used the scalar equations in [42, 43] to study the effects of a surface mass layer on the resonance frequencies of a quartz resonator. A perturbation analysis is performed. The zero-order problem represents a resonator without the surface mass layer. The first-order problem results in a frequency perturbation integral which can be used to calculate the frequency shift caused by the mass layer. Simple and useful examples for the application of the perturbation integral are presented.

## II. SCALAR DIFFERENTIAL EQUATION

Consider a partially electroded quartz plate of thickness $2h$ and mass density $\rho$ as shown in Fig. 1. It is bounded by two parallel planes at $x_2 = \pm h$. The $x_2$ axis is normal to the plate. $x_1$ and $x_3$ are in the middle plane of the plate. The upper surface of the plate is covered with a thin mass layer whose mechanical effect is described by an impedance function $Z$ in time-harmonic motions. For an AT-cut quartz plate in time-harmonic free vibration at a frequency $\omega$, the in-plane variation of the $m$th-order thickness-shear mode is described by $u_1^m(x_1, x_3)$ which is governed by the following scalar equation [42]:

$$M_m \frac{\partial^2 u_1^m}{\partial x_1^2} + c_{55} \frac{\partial^2 u_1^m}{\partial x_3^2} - \hat{\bar{c}}_{66}\left(\frac{m\pi}{2h}\right)^2 u_1^m + \rho\omega^2 u_1^m = \frac{i\omega}{h} Z(\omega) u_1^m, \quad (1)$$

where

$$m = 1, 3, 5, \cdots, \quad \hat{\bar{c}}_{66} = \begin{cases} \bar{c}_{66}, & \text{unelectroded,} \\ \hat{c}_{66}, & \text{electroded,} \end{cases}$$

$$M_m = c_{11} + (c_{12} + c_{66})r + \frac{4(r\bar{c}_{66} - c_{66})(rc_{22} + c_{12})}{c_{22} m\pi\kappa} \cot\frac{\kappa m\pi}{2},$$

$$\bar{c}_{66} = c_{66} + \frac{e_{26}^2}{\varepsilon_{22}}, \quad \kappa = \left(\frac{\bar{c}_{66}}{c_{22}}\right)^{1/2}, \quad r = \frac{c_{12} + c_{66}}{\bar{c}_{66} - c_{22}}, \quad (2)$$

$$\hat{c}_{66} = \bar{c}_{66}\left(1 - \frac{8\bar{k}_{26}^2}{n^2\pi^2} - 2R\right), \quad \bar{k}_{26}^2 = \frac{e_{26}^2}{\bar{c}_{66}\varepsilon_{22}}, \quad R = \frac{2\rho' h'}{\rho h}.$$

The equation is slightly different in electroded and unelectroded regions as indicated by the different values of $\hat{\bar{c}}_{66}$ in (2). $\rho'$ and $2h'$ are the density and thickness of the electrodes. $c_{pq}$, $e_{ip}$ and $\varepsilon_{ij}$ are the usual elastic, piezoelectric and dielectric constants. The impedance $Z$ in (2) is due to the surface mass layer and it may depend on $\omega$, $x_1$ and $x_3$ in general. The continuity conditions on $C_1$ between the electroded and unelectroded areas as well as the boundary conditions on $C_2$ are

$$[\![u_1^m]\!] = 0 \quad \text{and} \quad \left[\!\!\left[\frac{\partial u_1^m}{\partial n}\right]\!\!\right] = 0 \quad \text{on} \quad C_1,$$

$$u_1^m = 0 \quad \text{or} \quad \frac{\partial u_1^m}{\partial n} = 0 \quad \text{on} \quad C_2. \quad (3)$$

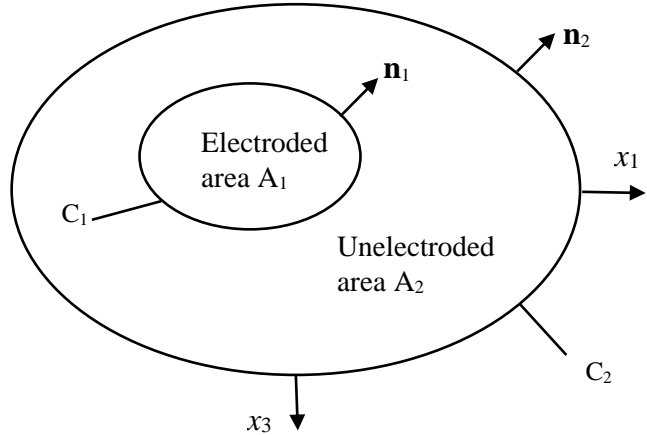

Fig. 1. A partially electroded quartz plate and coordinate system. Total area $A = A_1 + A_2$.

For convenience we denote

$$D_{11} = M_m, \quad D_{13} = D_{31} = 0, \quad D_{33} = c_{55}. \quad (4)$$

We also introduce a 2-D summation convention that indices *a* and *b* assume 1 and 3 only but not 2. Then (1) takes the following form which is also valid for doubly-rotated quartz plates when proper $D_{ab}$ are used:

$$D_{ab}u^m_{1,ab} - \hat{\bar{c}}_{66}\left(\frac{m\pi}{2h}\right)^2 u^m_1 + \rho\omega^2 u^m_1 = \frac{i\omega}{h}Z(\omega)u^m_1. \tag{5}$$

### III. PERTURBATION ANALYSIS

We are interested in the case of a thin mass layer with a small impedance. Its effect on the frequency is also small. Let

$$\omega = \bar{\omega} + \Delta\omega,$$
$$u^m_1 = \bar{u} + \Delta u, \tag{6}$$

which $\bar{\omega}$ and $\bar{u}$ are the frequency and mode when the mass layer is not present ($Z = 0$). $\Delta\omega$ and $\Delta u$ are also small. Substituting (6) into (5) and (3), for small $\Delta\omega$, $\Delta u$ and $Z$, we collect their zero- and first-order terms separately to obtain

$$D_{ab}\bar{u}_{,ab} - \hat{\bar{c}}_{66}\left(\frac{m\pi}{2h}\right)^2 \bar{u} + \rho\bar{\omega}^2 \bar{u} = 0, \tag{7}$$

$$\bar{u} = 0 \quad \text{and} \quad \left\|\frac{\partial \bar{u}}{\partial n}\right\| = 0 \quad \text{on} \quad C_1,$$
$$\bar{u} = 0 \quad \text{or} \quad \frac{\partial \bar{u}}{\partial n} = 0 \quad \text{on} \quad C_2, \tag{8}$$

and

$$D_{ab}(\Delta u)_{,ab} - \hat{\bar{c}}_{66}\left(\frac{m\pi}{2h}\right)^2 (\Delta u)$$
$$+ \rho 2\bar{\omega}(\Delta\omega)\bar{u} + \rho\bar{\omega}^2 (\Delta u) = \frac{i\bar{\omega}}{h}Z(\bar{\omega})\bar{u}, \tag{9}$$

$$\Delta u = 0 \quad \text{and} \quad \left\|\frac{\partial(\Delta u)}{\partial n}\right\| = 0 \quad \text{on} \quad C_1,$$
$$\Delta u = 0 \quad \text{or} \quad \frac{\partial(\Delta u)}{\partial n} = 0 \quad \text{on} \quad C_2. \tag{10}$$

The zero-order problem represents the case when the mass layer is not present. Its solutions $\bar{\omega}$ and $\bar{u}$ are assumed known. We want to determine $\Delta\omega$ from the first-order problem governed by (9) and (10). Multiplying (9) by $\bar{u}$, integrating the resulting equation over A = $A_1$ + $A_2$, using the 2-D divergence theorem twice, employing the continuity and boundary conditions in (8) and (10), we obtain

$$\int_A \left\{D_{ab}\bar{u}_{,ab} - \hat{\bar{c}}_{66}\left(\frac{m\pi}{2h}\right)^2 \bar{u} + \rho\bar{\omega}^2\bar{u}\right\}(\Delta u)dA$$
$$+ \int_A \rho 2\bar{\omega}(\Delta\omega)\bar{u}\bar{u}dA = \int_A \frac{i\bar{\omega}}{h}Z(\bar{\omega})\bar{u}\bar{u}dA. \tag{11}$$

The left-hand side of (11) vanishes because of (7). Then (11) leads to the following frequency perturbation integral:

$$\Delta\omega = \frac{i}{2h} \frac{\int_A Z(\bar{\omega}) \bar{u} \bar{u} dA}{\int_A \rho \bar{u} \bar{u} dA}. \quad (12)$$

## IV. APPLICATIONS

Consider the simple case of an unbounded and unelectroded AT-cut quartz plate with a mass layer on its top surface. The relevant acoustic impedance of the layer is defined by [42]

$$T_{21} = -Z\dot{u}_1, \quad x_2 = h. \quad (13)$$

When the mass layer thickness is $2h^l$ and its mass density is $\rho^l$, the impedance is given by [42]

$$Z(\omega) = i\omega 2\rho^l h^l. \quad (14)$$

In this case the frequencies and modes when the mass layer is not present is given by

$$\bar{\omega} = \frac{m\pi}{2h}\sqrt{\frac{c_{66}}{\rho}}, \quad \bar{u} = 1. \quad (15)$$

For a uniform mass layer with constant $h^l$ and $\rho^l$, the perturbation integral in (12) leads to

$$\frac{\Delta\omega}{\bar{\omega}} = -\frac{1}{h}\frac{\int_A \rho^l h^l \bar{u}\bar{u} dA}{\int_A \rho \bar{u}\bar{u} dA} = -\frac{\rho^l h^l}{\rho h}, \quad (16)$$

which is the well-known Sauerbrey equation [44] for the mass sensitivity of QCMs. This serves as a verification of (12).

In real QCM applications a nonuniform mass layer is a common situation [45] for which the Sauerbrey equation is no longer valid. As an example of a nonuniform mass layer, consider

$$h^l = h^0 \exp[-\lambda(x_1^2 + x_3^2)] = h^0 \exp[-\lambda r^2]. \quad (17)$$

In this case the perturbation integral in (12) gives

$$\frac{\Delta\omega}{\bar{\omega}} = -\frac{\rho^l}{h\rho}\frac{\int_A h^0 \exp(-\lambda r^2) \bar{u}\bar{u} dA}{\int_A \bar{u}\bar{u} dA} = -\frac{\rho^l}{h\rho}\frac{\int_0^R h^0 \exp(-\lambda r^2) 2\pi r dr}{\int_0^R 2\pi r dr}$$

$$= -\frac{\rho^l h^0}{\rho h}\frac{1-\exp(-\lambda R^2)}{\lambda R^2}, \quad (18)$$

When $\lambda = 0$, (18) reduces to the Sauerbrey equation in (16). When $\lambda > 0$, (17) describes a layer thicker in the middle and thinner near its edge. In most applications the mass layer thickness varies slowly with a small $\lambda$. As $\lambda$ increases from zero, when $\lambda$ is still small, from (18) we have

$$\frac{\Delta\omega}{\bar{\omega}} \cong -\frac{\rho^l h^0}{\rho h}\frac{1-(1-\lambda R^2 + \lambda^2 R^4/2)}{\lambda R^2} = -\frac{\rho^l h^0}{\rho h}\left(1 - \frac{\lambda R^2}{2}\right). \quad (19)$$

For a nonzero and positive value of $\lambda$, the mass layer is thinner near its edge. In this case (19) shows that there is less frequency drop because there is less inertia near the plate edge. The large-$\lambda$ limit of (18) is given by

$$\frac{\Delta\omega}{\bar{\omega}} \cong -\frac{\rho^l h^0}{\rho h}\frac{1}{\lambda R^2}. \quad (20)$$

In general (18) can be written as

$$\frac{\Delta\omega}{\bar{\omega}} = -\frac{\rho^l h^0}{\rho h} f(\lambda R^2), \quad (21)$$

where
$$f(x) = \frac{1-\exp(-x)}{x}. \qquad (22)$$

$f$ versus $x$ is shown in Fig. 2. It is a monotonically decreasing function irrespective of the magnitude of $\lambda$.

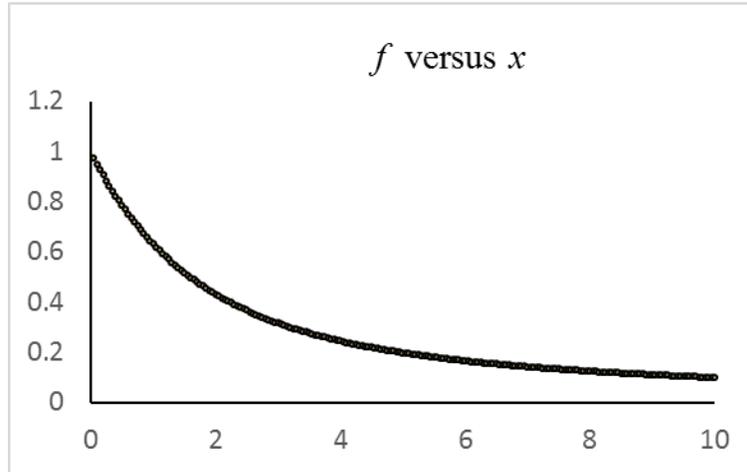

Fig. 2. $f(x)$.

## V. CONCLUSIONS

A first-order perturbation integral is obtained for the frequency shift induced by a thin mass layer on the top of a quartz thickness-shear acoustic wave resonator. In the special case of an unbounded and unelectroded plate with a uniform mass layer, the perturbation integral reduces to the well-known Sauerbrey equation. The basic effect of a nonuniform mass layer is shown by an example. The results are of fundamental importance and usefulness in the understanding and design of QCMs.

## REFERENCES
[1] H. F. Tiersten, "Analysis of intermodulation in thickness-shear and trapped energy resonators," *J. Acoust. Soc. Am.*, vol. 57, no. 3, pp. 667-681, 1975.
[2] H. F. Tiersten, "Analysis of trapped-energy resonators operating in overtones of coupled thickness shear and thickness twist," *J. Acoust. Soc. Am.*, vol. 59, no. 4, pp. 879-888, 1976.
[3] H.F. Tiersten, "Analysis of overtone modes in monolithic crystal filters," *J. Acoust. Soc. Am.*, vol. 62, no. 6, pp. 1424-1430, 1977.
[4] H. F. Tiersten and R. C. Smythe, "An analysis of contoured crystal resonators operating in overtones of coupled thickness shear and thickness twist," *J. Acoust. Soc. of Am.*, vol. 65, no. 6, pp. 1455-1460, 1979.
[5] D. S. Stevens and H. F. Tiersten, "An analysis of doubly rotated quartz resonators utilizing essentially thickness modes with transverse variation," *J. Acoust. Soc. Am.*, vol. 79, no. 6, pp. 1811-1826, 1986
[6] B. K. Sinha, "Doubly rotated contoured quartz resonator," *IEEE Trans. Ultrason., Ferroelect., Freq. Contr.*, vol. 48, no. 5, pp. 1162-1180, 2001.

[7] E. P. EerNisse, "Analysis of thickness modes of contoured doubly rotated, quartz resonators," *IEEE Trans. Ultrason., Ferroelect., Freq. Contr.*, vol. 48, no. 5, pp. 1351-1361, 2001.

[8] B. K. Sinha and D. S. Stevens, "Thickness-shear vibrations of a beveled AT-cut quartz plate," *J. Acoust. Soc. Am.*, vol. 66, no. 1, pp. 192-196, 1979.

[9] A. V. Apostolov and S. H. Slavov, "Frequency spectrum and modes of vibration in circular, convex AT-cut beveled-design quartz resonators," *Applied Physics A*, vol. 29, no. 1, pp. 33-37, 1982.

[10] S. Hertl, L. Wimmer and E. Benes, "Investigation of the amplitude distribution of AT-cut quartz crystals," *J. Acoust. Soc. Am.*, vol. 78, no. 4, pp. 1337-1343, 1985.

[11] H. F. Tiersten and R. C. Smythe, "Coupled thickness-shear and thickness-twist vibrations of unelectroded AT-cut quartz plates," *J. Acoust. Soc. Am.*, vol. 78, no. 5, pp. 1684-1689, 1985.

[12] S. H. Slavov, "Modes of vibration, motion inductance, and resonance interval of circular, convex AT-cut beveled design trapped energy quartz resonators," *Applied Physics A*, vol. 40, no. 1, pp. 59-65, 1986.

[13] S. H. Slavov, "Equivalent resonance radius of contoured AT-cut quartz resonators," *Appl. Phys. A*, vol. 43, pp. 111-116, 1987.

[14] S. H. Slavov and D. G. Ouroushev, "Investigation of harmonic modes of vibration in spherically contoured AT-cut quartz resonators by the degenerate hypergeometric function," *J. Phys. D: Appl. Phys.*, vol. 23, pp. 434-438, 1990.

[15] E. P. EerNisse, L. D. Clayton and M. H. Watts, "Distortions of thickness shear mode shapes in plano-convex quartz resonators with mass perturbations," *IEEE Trans. Ultrason., Ferroelect., Freq. Contr.*, vol. 37, no. 6, pp. 571-576, 1990.

[16] E. Benes, M. Schmid and V. Kravchenko, "Vibration modes of mass-loaded planoconvex quartz crystal resonators," *J. Acoust. Soc. Am.*, vol. 90, no. 2, pp. 700-706, 1991.

[17] H. F. Tiersten and Y. S. Zhou, "Transversely varying thickness modes in quartz resonators with beveled cylindrical edges," *J. Appl. Phys.*, vol. 76, no. 11, pp. 7201-7208, 1994.

[18] H. F. Tiersten, B. J. Lwo and B. Dulmet, "Transversely varying thickness modes in trapped energy resonators with shallow and beveled contours," *J. Appl. Phys.*, vol. 80, no. 2, pp. 1037-1046, 1996.

[19] O. Shmaliy, "Eigenvibrations in trapped-energy contoured piezoelectric resonators with a one-sided arbitrarily oriented elliptical convexity," *International Journal of Solids and Structures*, vol. 43, no. 25-26, pp. 7869-7879, 2006.

[20] J. S. Yang, "An analysis of partially electroded, contoured quartz resonators with beveled cylindrical edges," *IEEE Trans. Ultrason., Ferroelect., Freq. Contr.*, vol. 54, no. 11, pp. 2407-2409, 2007.

[21] Z. T. Yang, J. S. Yang and Y. T. Hu, "Optimal electrode shape and size of doubly rotated quartz plate thickness mode piezoelectric resonators," *Appl. Phys. Lett.*, vol. 92, art. no. 103516, 2008.

[22] W. Zhang, Z. T. Yang and J. S. Yang, "Membrane analogy of the Stevens-Tiersten equation for essentially thickness modes in plate quartz resonators," *IEEE Trans. Ultrason., Ferroelect., Freq. Contr.*, vol. 55, no. 7, pp. 1665-1668, 2008.

[23] P. Li, F. Jin and J. S. Yang, "Thickness-shear vibration of an AT-cut quartz resonator with a hyperbolic contour," *IEEE Trans. Ultrason., Ferroelect., Freq. Contr.*, vol. 59, no. 5, pp. 1006-1012, 2012.

[24] W. J. Wang, R. X. Wu, J. Wang, J. K. Du and J. S. Yang, "Thickness-shear modes of an elliptic, contoured AT-cut quartz resonator," *IEEE Trans. Ultrason., Ferroelect., Freq. Contr.*, vol. 60, no. 6, pp. 1192-1198, 2013.

[25] H. J. He, J. S. Yang and Q. Jiang, "Thickness-shear and thickness-twist vibrations of circular AT-cut quartz resonators," *Acta Mechanica Solida Sinica*, vol. 26, no. 3, pp. 245-254, 2013.

[26] J. J. Shi, C. Y. Fan, M. H. Zhao and J. S. Yang, "Variational formulation of the Stevens-Tiersten equation and application in the analysis of rectangular trapped-energy quartz resonators," *J. Acoust. Soc. Am.*, vol. 135, no. 1, pp. 175-181, 2014.

[27] J. J. Shi, C. Y. Fan, M. H. Zhao and J. S. Yang, "Trapped thickness-shear modes in a contoured, partially electroded AT-cut quartz resonator," *Eur. Phys. J. Appl. Phys.*, vol. 69, art. no. 10302, 2015.

[28] J. J. Shi, C. Y. Fan, M. H. Zhao and J. S. Yang, "Variational analysis of thickness-shear vibrations of a quartz piezoelectric plate with two pairs of electrodes as an acoustic wave filter," *Int. J. Appl. Electromagnetics and Mechanics*, vol. 47, pp. 951-961, 2015.

[29] H. Chen, J. Wang, J. K. Du and J. S. Yang, "Thickness-shear modes and energy trapping in a rectangular piezoelectric quartz resonator with partial electrodes," *Ferroelectric Letters Section*, vol. 42, no. 1-3, pp. 10-17, 2015.

[30] Z. N. Zhao, Z. H. Qian, B. Wang and J. S. Yang, "Analysis of thickness-shear and thickness-twist modes of AT-cut quartz acoustic wave resonator and filter," *Appl. Math. Mech.*, vol. 36, no. 12, pp. 1527-1538, 2015.

[31] Z. N. Zhao, Z. H. Qian, B. Wang and J. S. Yang, "Thickness-shear and thickness-twist modes in an AT-cut quartz acoustic wave filter," *Ultrasonics*, vol. 58, pp. 1-5, 2015.

[32] J. J. Shi, C. Y. Fan, M. H. Zhao and J. S. Yang, "Thickness-shear vibration characteristics of an AT-cut quartz resonator with rectangular ring electrodes," *Int. J. Appl. Electromagnetics and Mechanics*, vol. 51, pp. 1-10, 2016.

[33] J. J. Shi, C. Y. Fan, M. H. Zhao and J. S. Yang, "Energy trapping of thickness-shear modes in inverted-mesa AT-cut quartz piezoelectric resonators," *Ferroelectrics*, vol. 494, no. 1, pp. 157-169, 2016.

[34] Z. N. Zhao, Z. H. Qian and B. Wang, "Effects of unequal electrode pairs on an x-strip thickness-shear mode multi-channel quartz crystal microbalance," *Ultrasonics*, vol. 72, pp. 73-79, 2016.

[35] Z. N. Zhao, Z. H. Qian and B. Wang, "Thickness-shear vibration of a Z-strip AT-cut quartz crystal plate with nonuniform electrode pairs," *Ferroelectrics*, vol. 506, no. 1, pp. 48-62, 2017.

[36] L. L. Yuan, J. K. Du, J. Wang and J. S. Yang, "Thickness-shear and thickness-twist vibrations of rectangular quartz crystal plates with nonuniform thickness," *Mechanics of Advanced Materials and Structures*, vol. 24, no. 11, pp. 937-942, 2017.

[37] B. Wang, X. Y. Dai, X. T. Zhao and Z. H. Qian, "A semi-analytical solution for the thickness-vibration of centrally partially-electroded circular AT-cut quartz resonators," *Sensors*, vol. 17, no. 8, 1820, 2017.

[38] J. J. Shi, C. Y. Fan, M. H. Zhao and J. S. Yang, "Effects of electrode off center on trapped thickness-shear modes in contoured AT-cut quartz resonators," *International Journal of Acoustics and Vibration*, vol. 23, no. 4, pp. 423-431, 2018.

[39] L. T. Xie, S. Y. Wang, C. Z. Zhang and J. Wang, "An analysis of the thickness vibration of an unelectroded doubly-rotated quartz circular plate," *J. Acoust. Soc. Am.*, vol. 144, pp. 814-821, 2018.

[40] F. Zhu, B. Wang, X. Y. Dai, Z. H. Qian, I. Kuznetsova, V. Kolesov and B. Huang, "Vibration optimization of an infinite circular AT-cut quartz resonator with ring electrodes," *Applied Mathematical Modelling*, vol. 72, pp. 217-229, 2019.

[41] F. Zhu, P. Li, X.Y. Dai, Z. H. Qian, I. Kuznetsova, V. Kolesov and T. F. Ma, "A theoretical model for analyzing the thickness-shear vibration of a circular quartz crystal plate with

multiple concentric ring electrodes," *IEEE Trans. Ultrason., Ferroelect., Freq. Contr.*, vol. 68, pp. 1808-1818, 2021.

[42] H. J. He, J. S. Yang and J. A. Kosinski, "Scalar differential equation for slowly-varying thickness-shear modes in AT-cut quartz resonators with surface impedance for acoustic wave sensor application," *IEEE Sensors Journal*, vol. 13, no. 11, pp. 4349-4355, 2013.

[43] H. J. He, J. S. Yang, J. A. Kosinski and H. F. Zhang, "Scalar differential equations for transversely varying thickness modes in doubly-rotated quartz crystal sensors," *IEEE Sensors Letters*, vol. 2, no. 3, art. no. 2501004, 2018.

[44] G. Sauerbrey, "Verwendung von schwingquarzen zur wägung dünner schichten und zur mikrowägung," *Zeitschrift für Physik*, vol. 155, pp. 206-222, 1959.

[45] J. R. Vig and A. Ballato, "Comments about the effects of nonuniform mass loading on a quartz crystal microbalance," *IEEE Trans. Ultrason., Ferroelect., Freq. Contr.*, vol. 45, no. 5, pp. 1123-1124, 1998.